\documentclass{article}
\usepackage[T1]{fontenc}
\usepackage[utf8]{inputenc}
\usepackage{ismir} 
\usepackage{amsmath,cite,url}
\usepackage{graphicx}
\usepackage{color}
\usepackage{booktabs}
\usepackage{multirow}
\usepackage{makecell}
\usepackage{arydshln}

\usepackage{xcolor}

\usepackage{tablefootnote, colortbl, multirow}

\usepackage[
  activate={true,nocompatibility},
  final,
  tracking=true,
  kerning=true,
  spacing=true,
  protrusion=true,
  expansion=true,
  factor=1100,
  stretch=30,
  shrink=30
]{microtype}
\microtypecontext{spacing=nonfrench}

\usepackage{flushend}

\title{Assessing the Alignment of Audio Representations with Timbre Similarity Ratings}

\multauthor
  {Haokun Tian$^1$ \hspace{1cm} Stefan Lattner$^2$ \hspace{1cm} Charalampos Saitis$^1$}
  {$^1$ Queen Mary University of London, UK \hspace{1cm} $^2$ Sony Computer Science Laboratories, Paris, France\\
  {\tt\small haokun.tian@qmul.ac.uk}
  }

\def\authorname{H. Tian, S. Lattner, and C. Saitis}

\usepackage[bookmarks=false,pdfauthor={\authorname},pdfsubject={\pdfsubject},hidelinks]{hyperref}

\begin{document}

\maketitle

\begin{abstract}
Psychoacoustical so-called ``timbre spaces'' map perceptual similarity ratings of instrument sounds onto low-dimensional embeddings via multidimensional scaling, but suffer from scalability issues and are incapable of generalization. Recent results from audio (music and speech) quality assessment as well as image similarity have shown that deep learning is able to produce embeddings that align well with human perception while being largely free from these constraints. Although the existing human-rated timbre similarity data is not large enough to train deep neural networks (2,614 pairwise ratings on 334 audio samples), it can serve as test-only data for audio models.
In this paper, we introduce metrics to assess the alignment of diverse audio representations with human judgments of timbre similarity by comparing both the absolute values and the rankings of embedding distances to human similarity ratings. Our evaluation involves three signal-processing-based representations, twelve representations extracted from pre-trained models, and three representations extracted from a novel sound matching model. Among them, the style embeddings inspired by image style transfer, extracted from the CLAP model and the sound matching model, remarkably outperform the others, showing their potential in modeling timbre similarity.
\end{abstract}

\section{Introduction}\label{sec:introduction}

\begin{figure}[ht]
    \centering
    \includegraphics[width=0.45\textwidth]{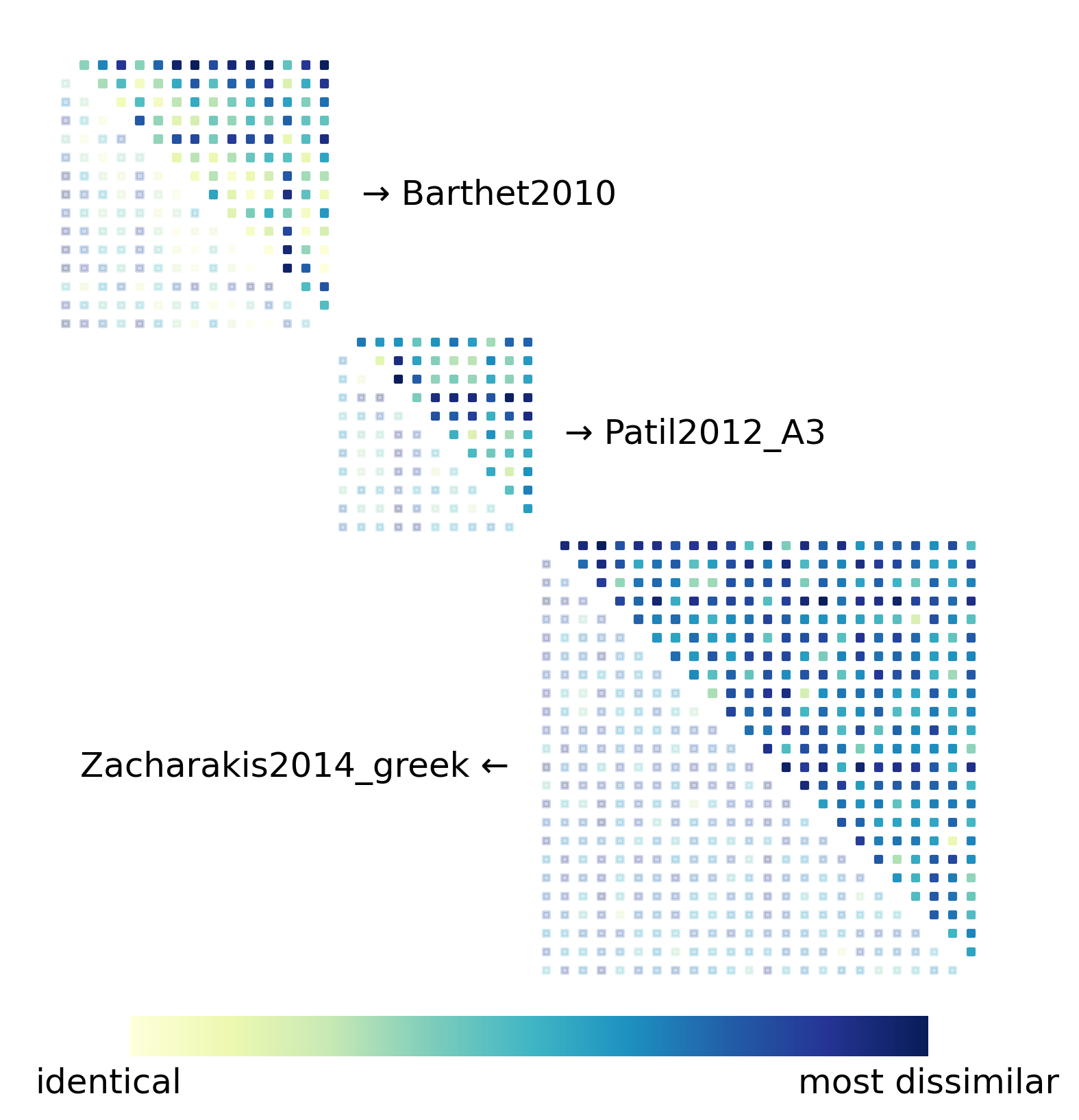}
    \caption{Human similarity ratings of three timbre space datasets \cite{barthet2010clarinet, patil2012music, zacharakis2015interlanguage}. Each block matrix represents pairwise comparisons between audio stimuli. Entry $(i,j)$ indicates the perceived similarity between audio $i$ and audio $j$. Darker colors indicate lower similarity. The lower triangular part is made partially transparent due to symmetry.}
    \label{fig:true_dissim}
\end{figure}

How do humans distinguish between different musical timbres? This question has driven research in the field of psychoacoustics for decades \cite{mcadams2019perceptual}. Researchers typically recruit a group of people, play different sounds to them in a controlled acoustic environment, and ask them to rate the differences in the sounds by assigning a score. These sounds are normalized in pitch, loudness, and duration so that participants can focus on timbre. After collecting all human ratings, researchers use a technique called multidimensional scaling (MDS) to map the sounds onto a low-dimensional space called \textit{timbre space}, where distances between the resulting embeddings reflect human ratings. This space is then analyzed to find whether certain acoustic features can explain how timbres are ordered along different dimensions. 

However, these studies have always been limited in scale—that is, they only involved a small number of sound stimuli. If the number of sounds were to increase, the required human ratings would grow quadratically, since ratings are given in pairs. The largest study to date is \cite{elliott2013acoustic}, which includes 42 sound stimuli.
Additionally, pitch, loudness, and duration must be fixed to eliminate confounding effects, meaning that a timbre space with variable pitch or loudness can never be constructed. Furthermore, to place new audio samples into a timbre space, new human ratings must always be collected, indicating that the timbre space as a model does not have the ability to generalize without additional efforts.

To overcome these limitations, we envision a neural network-based model that can serve as a perceptual metric for timbre—that is, it can embed any audio input into a space where distances between embeddings reflect how different humans perceive them. Such a metric would have numerous applications. First, to train a model to generate instrument sounds \cite{engel2017neural, engel2020ddsp}, one could match the timbre by directly computing the loss using this metric rather than relying on spectrogram distances.
In music production, it could enable efficient retrieval of samples with similar or contrasting timbres without the need for laborious listening. Additionally, it can encode audio signals into meaningful timbre tokens for generative modeling. Finally, it could provide a control space for musical expression, as has already been demonstrated \cite{wessel1979timbre}.

In this work, we take the first step by evaluating a range of models' representations, including signal-processing-based low-level representations and those produced by pre-trained audio models, on the past 21 small datasets collected by psychoacoustical timbre space studies. Our goal is to determine which model currently performs best in terms of matching human ratings of timbre similarity. We also include a newly trained sound matching model that predicts synthesis parameters from audio, and we evaluate audio representations extracted from it. In particular, we have found that style embeddings, originally proposed for image style transfer, are promising as a future direction for modeling timbre similarity. We clarify that the term ``alignment'' in this paper does not refer to any temporal synchronizations, and our assessment does not indicate the extent to which models produce disentangled timbre representations—an ability required for some of the applications mentioned in the previous paragraph.

\section{Related Work}\label{sec:relatedwork}

Several works have aimed to train models that produce a perceptual timbre space. Esling et al. \cite{esling2018generative} trained a variational autoencoder to reconstruct audio samples of different timbres, using perceptual ratings from timbre space studies to regularize the space. Lostanlen et al. \cite{lostanlen2021time} collected timbre similarity judgments on 78 sounds using free sorting \cite{chollet2014free}. By aggregating responses from different participants, the sounds were assigned to 19 clusters according to their perceptual similarity. They, along with a subsequent study \cite{vahidi2023perceptual}, developed metric learning models to capture this structure. Thoret et al. \cite{thoret2021learning} and Pascal et al. \cite{pascal2024robustness} attempted to learn distance metrics directly from timbre space data but did not learn a single metric that generalizes across all datasets. The most similar work to ours is by Vahidi et al. \cite{vahidiacoustic}, which evaluated the alignment of three representations with timbre similarity ratings. Additionally, the accompanied codebase \cite{hayes2021timbre} provides more implementations of alignment scores, but no results have been reported. Our work is a direct continuation, providing a holistic evaluation of various models and implementations with extended functionalities.

In broader domains, perceptual metrics have been developed to handle various levels of perturbations. They are either trained from scratch or fine-tuned using human perceptual judgments. For audio, Manocha et al. proposed DPAM \cite{manocha2020differentiable} and CDPAM \cite{manocha2021cdpam}, which address low-level audio perturbations such as noise addition, reverb, and compression. In the image domain, Zhang et al. \cite{zhang2018unreasonable} modeled low-level perturbations including traditional distortions such as noise addition and blur, as well as CNN-based distortions. Fu et al. \cite{fu2023dreamsim} explored mid-level perturbations related to pose, color, and shape. Muttenthaler et al. \cite{muttenthaler2023improving} investigated high-level perturbations involving different object-level concepts (e.g., ``grass'' versus ``sand''). The smallest dataset used is NIGHTS \cite{fu2023dreamsim} with 20K human judgments, while the largest is THINGS \cite{hebart2023things}, which contains 4.7M human judgments. In particular, \cite{zhang2018unreasonable} demonstrated that deep learning representations, regardless of the training data, model, or task used, exhibit a promising ability to capture perceptual image similarity. Therefore, although current timbre space datasets provide only 2.6K pairwise judgments and are insufficient for training, they are worth being used for model evaluation.

\section{Data and Experiments}

\subsection{Data}

\begin{table*}[t!]
\centering
\resizebox{\textwidth}{!}{%
\renewcommand{\arraystretch}{0.9}
\begin{tabular}{@{}llcccclc@{}}
\toprule
\textbf{Study} & \textbf{Dataset} & \textbf{\makecell{$N$ of \\sounds}}  & \textbf{Pitch} & \textbf{\makecell{Length \\(s)}}  & \textbf{\makecell{Loudness \\(dB LUFS)}}  & \textbf{Type of sounds} & \textbf{\makecell{$N$ of \\raters$^\ast$}} \\ 
\midrule
Grey (1977) \cite{grey1977multidimensional} & -- & 16  & E$\flat$4 & 0.27 $\pm$ 0.03 & -14.61 $\pm$ 1.57 & Synthesized from acoustic & 22 \\ 
\arrayrulecolor{white}\midrule
Grey \& Gordon (1978) \cite{grey1978perceptual} & -- & 16   & E$\flat$4 & 0.27 $\pm$ 0.03 & -14.87 $\pm$ 1.76 & Same as above, spectral envelopes traded & 22 \\ 
\midrule
Iverson \& Krumhansl (1993) \cite{iverson1993isolating} & Whole & 16  &  C4 & 3.19 $\pm$ 0.69 & -16.62 $\pm$ 3.99 & Acoustic \cite{opolko1987cd} & 10 \\ 
 & Onset & 16  & C4 & 0.11 $\pm$ 0.01 & -23.24 $\pm$ 9.10 & Same as above, only $\sim$80ms attack & 9 \\ 
 & Remainder & 16  & C4 & 3.10 $\pm$ 0.70 & -16.71 $\pm$ 4.13 & Same as above, $\sim$80ms attack removed & 9 \\ 
\midrule
McAdams et al.~(1995) \cite{mcadams1995perceptual} & -- & 18  & E$\flat$4 & 0.69 $\pm$ 0.19 & -18.36 $\pm$ 4.05 & FM simulated and hybrids & 22 \\ 
\midrule
Lakatos (2000) \cite{lakatos2000common} & Combined & 20  & E$\flat$4 & 1.50 & -24.08 $\pm$ 2.98 & Acoustic \cite{opolko1987cd} & 34 \\
        & Harmonic & 17  & E$\flat$4 & 1.50 & -25.03 $\pm$ 2.36 & Same as above, only harmonic sounds & 34 \\ 
        & Percussive & 18  & E$\flat$4 & 1.49 $\pm$ 0.05 & -23.97 $\pm$ 3.83 & Same as above, only percussive sounds & 34 \\ 

\midrule
Barthet et al.~(2010) \cite{barthet2010clarinet} & -- & 15 & E$\flat$4 & 1.56 $\pm$ 0.04 & -23.77 $\pm$ 1.13 & Physical modeling, clarinet only  & 16 \\ 
\midrule
Patil et al.~(2012) \cite{patil2012music} & A3 & 11 & A3 & 0.25 & -19.03 $\pm$ 0.85 &  Acoustic \cite{goto2003rwc} & 6 \\ 
            & D4 & 11 & D4 & 0.25 & -19.17 $\pm$ 0.81 & Same as above & 20  \\ 
            & G$\sharp$4 & 11  & G$\sharp$4 & 0.25 & -18.97 $\pm$ 0.81 & Same as above & 6 \\ 
\midrule
Zacharakis et al.~(2015) \cite{zacharakis2015interlanguage}  & Greek raters & 24  & A1--4 & 1.30 & -29.36 $\pm$ 3.84 & Acoustic \cite{opolko2006dvd} and synthesizers & 33 \\ 
  & English raters & 24  & A1--4 & 1.30 & -29.36 $\pm$ 3.84 & Same as above & 20 \\ 
\midrule
Siedenburg et al.~(2016) \cite{siedenburg2016acoustic}   & Exp 2A Set 1 & 14  & E$\flat$4 & 0.50 & -23.56 $\pm$ 3.15 & Acoustic \cite{vienna_symphonic_library} & 24 \\ 
  & Exp 2A Set 2 & 14 & E$\flat$4 & 0.50 & -23.77 $\pm$ 1.90 & Chimeric (spectral envelopes traded) & 24 \\ 
                & Exp 2A Set 3 & 14  & E$\flat$4 & 0.50 & -23.21 $\pm$ 2.37 & Acoustic and chimeric & 24 \\ 
                & Exp 2B & 14 & E$\flat$4 & 0.50 & -23.21 $\pm$ 2.37 & Same as above & 24 \\ 
\midrule
Saitis \& Siedenburg (2020) \cite{saitis2020brightness}  & GEdissim & 14  & E$\flat$4 & 0.50 & -23.56 $\pm$ 3.15 & Same as Exp 2A Set 1 in \cite{siedenburg2016acoustic,vienna_symphonic_library}  & 40 \\ 
\midrule
Vahidi et al.~(2020) \cite{vahidi2020timbre} & -- & 15 & A4 & 1.00 & -9.73 $\pm$ 3.09 & Subtractive synthesis & 35 \\ 
\arrayrulecolor{black}\bottomrule\rule{0pt}{0ex}\\
\end{tabular}
}
\\\vspace{-6pt}
  \raggedright
  \footnotesize{\tiny $^\ast$ Raters are typically musicians or with some sort of musical background. Some studies used a mixture of musicians and non-musicians.}\\
  \centering
  \caption{Summary of the 21 timbre similarity datasets.  We computed integrated loudness for each sample using \texttt{pyloudnorm} \cite{steinmetz2021pyloudnorm} with a block size of 0.08 seconds—slightly shorter than the shortest sample. While some dynamic variations can be observed, loudness is typically reported to have been normalized by expert listeners.}
  \label{tab:datasets}
  \vspace{-0.5\baselineskip}
\end{table*}

We use data curated by Thoret et al.~\cite{thoret2021learning} and Vahidi et. al~\cite{vahidiacoustic},\footnote{https://github.com/ben-hayes/timbre-dissimilarity-metrics}
comprising a total of 21 datasets from 11 published psychoacoustic studies~\cite{barthet2010clarinet, grey1977multidimensional, grey1978perceptual, iverson1993isolating, lakatos2000common, mcadams1995perceptual, patil2012music, saitis2020brightness, siedenburg2016acoustic, vahidi2020timbre, zacharakis2015interlanguage}. 
We present summary information for each dataset in Table~\ref{tab:datasets}. Each dataset contains a set of audio samples along with pairwise timbre similarity ratings. The sounds span a wide range: acoustic recordings of musical instrument notes, digitally edited acoustic samples designed to create simpler or cross-instrument spectra, and tones from synthesizers and electromechanical instruments. Each similarity rating is an absolute value calculated by averaging across multiple human listeners. We compile all ratings from the 21 datasets into a large, sparse block-diagonal matrix, where the upper triangular part of each block represents the ratings for unique audio pairs in the corresponding dataset. Figure~\ref{fig:true_dissim} illustrates three of these blocks, with symmetric pairwise ratings visualized by color. In total, the matrix includes 334 audio samples and 2,614 similarity ratings.

\subsection{Metrics}

Our goal is to investigate whether an audio model can perceive timbre similarity in a way that aligns with human perception. To this end, we compute pairwise distances between audio representations produced by the model to estimate similarity ratings, which we then compare with real human ratings to obtain alignment scores. Conceptually, the pairwise distances of audio representations form a \textit{predicted} similarity matrix, which we compare against the \textit{ground truth} similarity matrix, consisting of human ratings. 

To conduct this evaluation, we first obtain equally shaped representations for audio of different lengths, in preparation for the distance computation. Second, we employ a distance function to quantify their similarity. Third, we require metrics to measure the alignment between the predicted and true similarity matrices. Below, we present our approach to these three steps, resulting in two methods for handling variable length (see next section for cases with only one viable option), two methods for computing distances, and five ways of computing alignment scores. This gives a total of 10 or 20 scores for each representation (one model can produce multiple representations). Additionally, we have developed this process into an easy-to-use Python package, as described in Section~\ref{sec:package}.

\subsubsection{Handling Variable Audio Length}\label{ref:handle_variable_length}
To produce representations for audio, models typically use an analysis window with a fixed length, which slides over time with a hop length to produce time-varying representations.
As shown in Table~\ref{tab:datasets}, some timbre space datasets have fixed audio lengths, while others do not. This creates the need to compute identically shaped representations for audio of different lengths, as models employing sliding windows inherently produce representations with a time dimension proportional to the input length. To address this, we provide two approaches. The first is to squash the time dimension by computing the average over it. The second, which we call \textit{dynamic-length}, matches input lengths by padding zeros to the right—or truncating in very few cases—to arrive at the same length. Below are a few scenarios. If the model has a long enough analysis window that covers all input lengths of the timbre space data, all audio samples are padded to this length. For this case, we will not compute the time-averaged version, as the representation has only one frame along the time dimension. And if the model operates with a rather short analysis window, the solution is to always dynamically pad the shorter audio within a pair to match the length of the longer one, ensuring the shapes of their representations match. We only truncate audio samples in one case, and that is when the model is sensitive to time shifting and does not accept sliding. We pad the shorter audio and truncate the longer audio to match the fixed window length. In this case we will also not compute the time-averaged version. We describe in Section~\ref{sec:other_models_and_dynamic_length_adaption} specifically how the \textit{dynamic-length} approach is applied for each model.

\subsubsection{Distance Functions} 
We compute pairwise distances using two functions: the $\ell_2$ distance and cosine distance (defined as one minus the cosine similarity). These functions are applied to flattened audio representations.

\subsubsection{Alignment Scores}
The alignment scores are computed between the predicted similarity matrix and the ground truth similarity matrix, both of which are diagonal block matrices. We note that only the pairs within the diagonal blocks are valid, have similarity ratings, and are used to compute the alignment scores. Pairs outside the diagonal blocks lack human ratings as samples are from different datasets. Before computation, the values in each dataset-level block matrix (from both the predicted and ground truth matrices) are rescaled to the range $[0,1]$.

We compute five alignment scores in total. The first is the mean absolute error (MAE). It computes the absolute error between ratings of each unique audio pair and averages the errors across all pairs. The other four are rank-based scores that evaluate how well the model ranks a list of audio samples based on their timbre similarity to a given sample. This is done by comparing each row of the two similarity matrices (column-wise comparison yields the same results due to symmetry), which contains the similarity ratings between a reference sample (whose index corresponds to the row index) and all other samples from the same dataset, where the relative ordering of these ratings determines the timbre similarity ranking with respect to the reference sample. Sorting the ratings per row yields $N$ rankings—one for each sample (each sample is used once as the reference sample)—where $N$ is the total number of samples. The predicted ranking from the computed matrix is then compared to the true ranking from the ground-truth matrix. The final rank-based alignment scores are obtained by computing a score per row and averaging across all rows. 

We use four rank-based metrics for per-row comparisons: Kendall’s rank correlation coefficient, Normalized Discounted Cumulative Gain (NDCG) that treats timbre similarity (rather than dissimilarity) as relevance, Spearman’s rank correlation coefficient, and triplet agreement with an adjustable margin. For the first three standard metrics, we use implementations from \texttt{torchmetrics} \cite{Nicki_2022}. For triplet agreement, we first extract all triplets from the ground-truth row that satisfies the given margin condition—that is, the absolute difference between two dissimilarity ratings must be greater than the margin value. This mimics a just-noticeable difference in human perception, where rating differences smaller than this threshold are considered perceptually equivalent to the reference and are therefore excluded from the evaluation. The margin is set to 0.1 in our evaluation. To illustrate, a triplet consists of three audio samples $(i,j,k)$, where $i$ is the reference sample and the similarity between $i$ and $j$ is compared to the similarity between $i$ and $k$. We then compute the triplet agreement rate as the ratio of triplets for which both matrix rows agree on the relative ordering of $j$ and $k$ with respect to $i$, to the total number of extracted triplets.

\subsubsection{The Python Package} \label{sec:package}

Our Python package\footnote{https://github.com/tiianhk/timbremetrics} \texttt{timbremetrics} simplifies the evaluation process by requiring only the input of a model capable of converting raw waveforms into audio representations to compute the alignment of that model with timbre similarity perception. This package supports all evaluations described in this section and additionally supports three distance functions: $\ell_1$, negative dot product, and Poincaré distance, a commonly used hyperbolic distance \cite{khrulkov2020hyperbolic}. It is compatible with audio models interfaced by the Fréchet Audio Distance Toolkit \cite{gui2024adapting}.\footnote{https://github.com/microsoft/fadtk} Furthermore, it can be used as a convenient training-time evaluation to inspect whether the model can acquire human-like ability to perceive timbre similarity during the learning process.

\subsection{Sound Matching}\label{sec:sound_matching}
We train a sound matching model that inverts the Vital synthesizer~\footnote{https://vital.audio} by predicting the synthesis parameters from synthesized audio.~\footnote{Code available at https://github.com/tiianhk/sm4tp} We are interested in whether, via this task, the model can learn meaningful representations that align with timbre similarity perception. We are motivated by the fact that synthesizer parameters encode all audio content using very little storage, and we expect that by learning this compression, we can obtain compact but expressive intermediate representations.

\subsubsection{Data Generation} We use \texttt{Vita},\footnote{https://github.com/DBraun/Vita}, a package that provides Python bindings for the Vital Synthesizer, to generate our dataset. Ten parameters and their ranges are subjectively selected to produce pronounced timbral changes when varied. These parameters include one discrete selection from seven basic waveshapes (e.g., sine wave, triangle wave) in the wavetable, two to distort the waveshape, two from a unison effect, three from the ADSR envelope, and two from an EQ. Of these, two are discrete parameters, and the remaining eight are continuous parameters. We uniformly sample each parameter to generate data. A random pitch is also sampled, but not as a prediction target, as pitch is considered unrelated to timbre. We generate a total of 500k samples, each with a maximum length of two seconds, which takes $\sim$8 hours on a single CPU core. All continuous parameters are rescaled to the range $[0, 1]$ to be used as regression targets.

\subsubsection{Model Architecture} The model starts with an initial convolutional layer with a large kernel size to capture low-level features, followed by batch normalization, ReLU activation, and max pooling. The core of the model comprises four residual blocks with increasing channel dimensions \cite{he2016deep}. Each residual block contains two convolutional layers with batch normalization and ReLU, along with a shortcut connection to preserve gradient flow. After feature extraction, a global average pooling layer reduces spatial dimensions, yielding a fixed-length embedding vector of size 256. This embedding is then fed into the output heads: a regression head that produces 8 continuous values between 0 and 1, and two classification heads for discrete predictions. Our model has $\sim$5M parameters.

\subsubsection{Training} For the discrete parameters, we compute cross-entropy losses. For the continuous parameters, we compute the $\ell_1$ losses. Losses are added without weighting. We use an 8/2 train-validation split, a batch size of 32, and the Adam optimizer with a learning rate of 1e-4. The model is trained for 100 epochs, which takes $\sim$12 hours on an A100 GPU. We observe that the validation loss for each parameter has converged.

\subsubsection{Task and Style Embeddings} We extract and evaluate three representations from the sound matching model. The first representation is the 256-dimensional embedding obtained right before the prediction heads, which encodes information necessary for solving the task—in our case, synthesis parameter prediction. We refer to this as the \textit{task embedding}. The other two are inspired by image style transfer \cite{gatys2016image, huang2017arbitrary}. We refer to them as the \textit{style embedding}.
Our motivation for using style embeddings is twofold: first, they are invariant to the spatial location of audio events on spectrograms; and second, they use features from different layers of a model, thus capturing multiple levels of abstraction—a property shown to be effective for modeling perceptual timbre similarity \cite{pascal2024robustness}.
The first style embedding, proposed by Gatys et al. \cite{gatys2016image}, is computed as a Gram matrix of feature activations, where each entry captures the inner product between two channels across the spatial dimensions, resulting in a symmetric matrix that captures correlations between channels. The second style embedding is shown to be effective by Huang and Belongie \cite{huang2017arbitrary} and is computed as the channel-wise mean and standard deviation of activations over the spatial dimensions. Given a feature map with shape $(B, C, H, W)$, where $B$ is the batch size, $C$ is the channel number, and $H$ and $W$ are the spatial dimensions, i.e. height and width, the style embedding of Gatys et al. has shape $(B, C, C)$, and the style embedding of Huang and Belongie has shape $(B, 2C,)$. With our model, we extract both Gatys style embeddings and Huang style embeddings from five intermediate convolutional layers: the initial convolutional layer, and the first convolutional layers in each of the four residual blocks. Feature maps are obtained after batch normalization. We concatenate the embeddings obtained from different layers, resulting in a single Gatys style embedding and a single Huang style embedding per input. Together with the tasking embedding, we abbreviate them as \textit{s.m.-task}, \textit{s.m.-Gatys}, and \textit{s.m.-Huang}.

\subsection{Other Models and Dynamic Length Adaption}\label{sec:other_models_and_dynamic_length_adaption}

We evaluate signal-processing-based methods\footnote{For audio at 44.1kHz, MFCC is computed with n\_mfcc=40, MSS is computed with fft\_sizes=(4096, 2048, 1024, 512, 256, 128), and JTFS is computed with $J=12$, $Q=(8,2)$, $J_{fr}=3$, and $Q_{fr}=2$ using the \texttt{Kymatio} package~\cite{andreux2020kymatio}.} that do not learn from data, including mel-frequency cepstral coefficients (MFCC), multi-scale spectrograms (MSS)~\cite{engel2020ddsp}, and the joint time-frequency scattering transform (JTFS)~\cite{anden2019joint}. We also evaluate pre-trained audio models interfaced through the Fréchet Audio Distance Toolkit~\cite{gui2024adapting}, which are typically trained on large datasets and used to evaluate generative models, as their representations are considered to correlate well with perceptual music quality. This includes three CLAP models trained with natural language supervision~\cite{CLAP2023, wu2023large}; two CDPAM models trained on perceptual ratings of audio quality~\cite{manocha2021cdpam}; and neural audio codecs including two Encodec models~\cite{defossez2022high} and one Descript Audio Codec (DAC)~\cite{kumar2023high}, which compress audio into lower-bitrate latent representations. Additionally, we evaluate Music2Latent~\cite{pasini2024music2latent} and a reproduction of the Complex Autoencoder (CAE)\cite{lattner2019learning} that produces representations invariant to transposition and time-shift.\footnote{Trained with piano music from the MAPS dataset~\cite{emiya2010maps}.}

\begin{figure}[ht]
    \centering
    \includegraphics[width=0.47\textwidth]{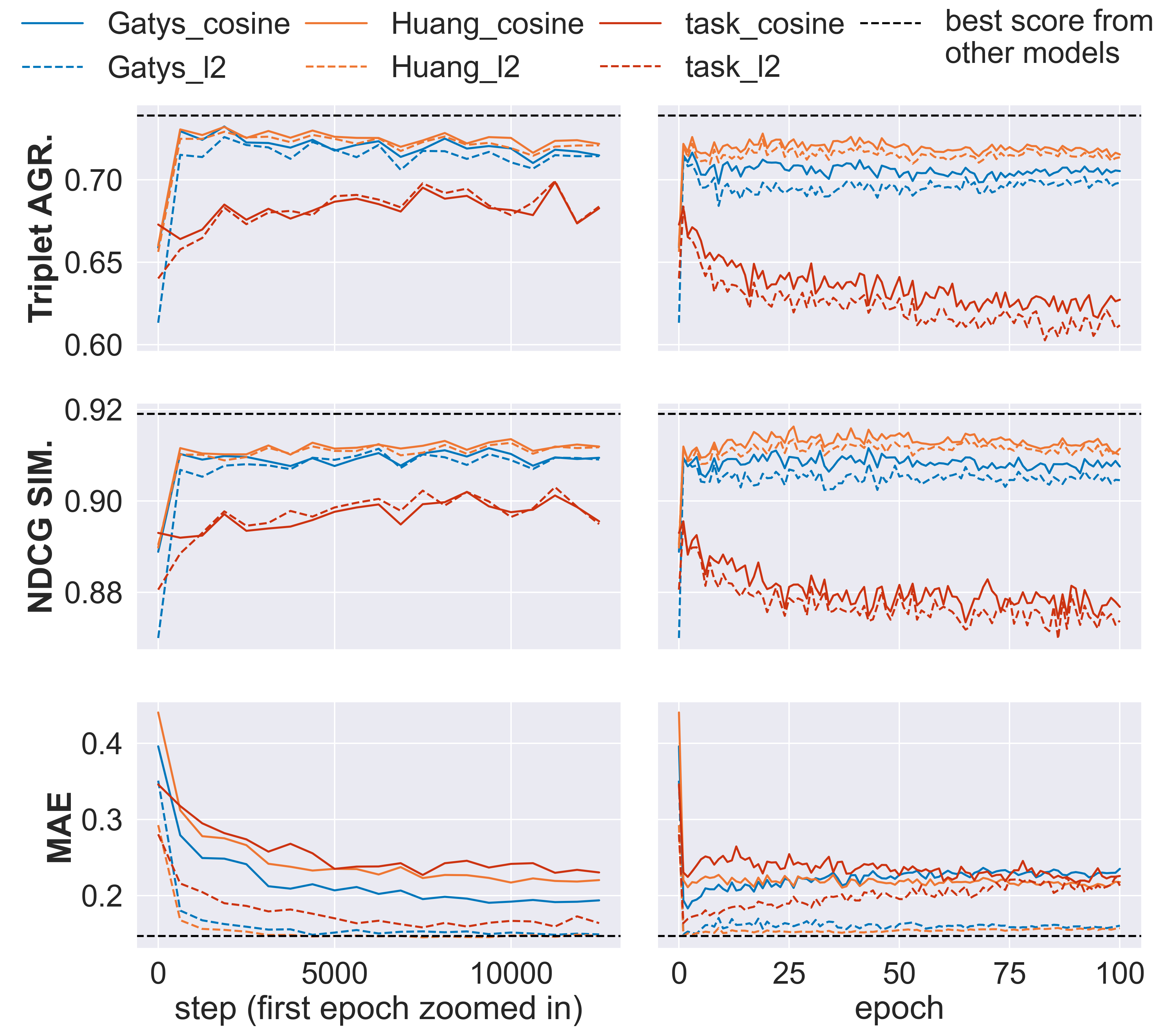}
    \caption{Evolution of alignment scores during the training of the sound matching model. The left column shows the detailed changes with the first epoch zoomed in, while the right column shows the overall progress across the total 100 epochs. For MAE, lower scores are better; for other metrics, higher scores are better. The best scores refer to the highest (or lowest, for MAE) values of each metric in Figure~\ref{fig:alignment_scores}.}
    \label{fig:sound_matching}
\end{figure}

\begin{figure*}[ht]
    \centering
    \includegraphics[width=1\textwidth]{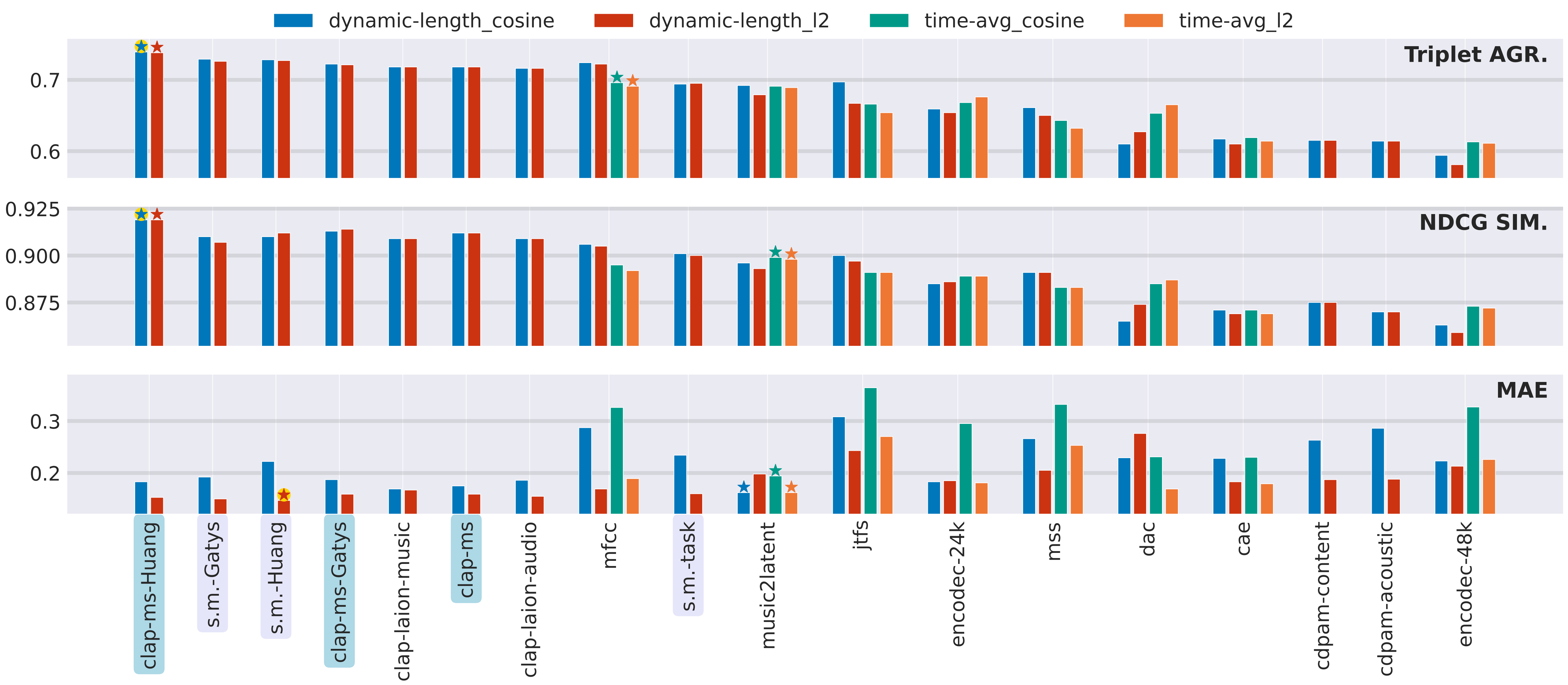}
    \caption{Evaluation results. Each configuration shown at the top is represented by a unique color. For each alignment score, the best result within each configuration is marked with a star, while a yellow dot behind the star indicates the overall winner. For MAE, lower scores are better; for other metrics, higher scores are better. Representations are ordered by the mean triplet agreement across configurations, in descending order. For text labels on the x-axis, the same background color indicates different representations extracted from the same model.}
    \label{fig:alignment_scores}
\end{figure*}

In Section~\ref{ref:handle_variable_length}, we discussed how to produce equally shaped representations for audio of different lengths. Here, we describe how this is specifically achieved for each model. Since the longest sample in our data is 4.39 seconds (see Table~\ref{tab:datasets}), we compute just one frame of representation for the following models, with their respective analysis window lengths indicated in parentheses: CLAP (ten seconds for~\cite{wu2023large} and seven seconds for~\cite{CLAP2023}) and CDPAM (five seconds). For these models, audio samples are padded to match the window length. The sound matching model has an analysis window of two seconds (which covers 19 out of 21 datasets) and is trained to capture the ADSR envelope, so the analysis window cannot be shifted. Therefore, we pad or truncate samples to a duration of two seconds. The above six models produce eight unique representations, for which we do not compute the time average since they contain only one time frame. For all other models, representations are computed using both time averaging and dynamic padding, where the shorter sample is padded to match the longer one within an audio pair.

We also evaluate style embeddings extracted from one CLAP model \cite{CLAP2023}, which differs from the sound matching model not only in training objective but also in architecture—it uses a Transformer backbone. However, style embeddings can be computed in a similar way: the internal representations produced by the Transformer consist of spatial tokens with a feature dimension, analogous to spatial locations and channel dimensions in CNNs. Therefore, we treat the transformer's feature dimension as equivalent to the channel dimension in CNNs. Following this analogy, we compute Gatys style embeddings by measuring correlations between feature dimensions, and Huang style embeddings by computing statistics (mean and variance) across each feature. We compute style embeddings using the outputs from each Swin Transformer block in the first three layers. Each layer contains multiple blocks, with nine blocks in total. The resulting embeddings are concatenated into a single embedding per input for evaluation.

\section{Results and Discussion}\label{sec:results}
In Figure~\ref{fig:sound_matching}~and~\ref{fig:alignment_scores}, we omit the Kendall and Spearman scores, as they are highly correlated with triplet agreement and produce nearly identical rankings of the evaluated representations. We choose to report triplet agreement instead, as it is a more intuitive metric to interpret than the other two.

Figure~\ref{fig:sound_matching} shows training-time alignment scores for our sound matching model on all 21 datasets. We use a three-fold validation-test split to ensure a fair comparison with other models shown in Figure~\ref{fig:alignment_scores}. For each score, we select the best-performing model based on validation performance using only checkpoints from the first epoch. Final results are reported as averages over the test folds. The style embeddings converge quickly, retaining high alignment with human timbre judgments, whereas the task embedding overfits and loses generalization over time. In Figure~\ref{fig:alignment_scores}, the style embeddings from both the sound matching model and the CLAP model show clear improvements over their base representations, demonstrating the effectiveness of style embeddings regardless of training objective or model architecture. In particular, the Huang style embedding extracted from the CLAP model shows the strongest performance.

MFCC remains competitive and outperforms many trained models. By contrast, CDPAM—trained on low-level distortion judgments of speech—does not adapt well to musical timbre. This may be due to differences in data domain, or it may echo findings from DreamSim~\cite{fu2023dreamsim}, which suggest that perceptual similarity learned for one type of perturbation does not generalize to others. MSS also underperforms, consistent with prior work~\cite{turian2020im} showing that spectral distances can be problematic for capturing perceptual similarity of pitch.

Interestingly, Encodec’s 24k model aligns better with human ratings than the 48k model, despite the former being trained on a variety of audio data, whereas the latter is trained exclusively on music. This suggests that increased compression combined with broader training data may encourage the model to discard irrelevant details while preserving perceptually meaningful structure, resulting in a more efficient internal timbre representation. This also highlights the effectiveness of Music2Latent, which aligns better with human ratings, has a high compression rate, and is trained on both music and speech. JTFS performs moderately on rank-based metrics but poorly on MAE. This suggests that its distance scale may be nonlinearly stretched relative to human perception, preserving the order of examples but distorting their absolute differences.

\section{Conclusion}
In this paper, we introduced a unified evaluation framework to compare model-derived distances with human similarity ratings from 21 classic timbre space datasets, encompassing a wide range of musical instrument sounds. We assessed both hand-crafted features (e.g., MFCC) and deep learning-based representations (e.g., CLAP, CDPAM, neural audio codecs), as well as a newly proposed sound matching model that inverts a wavetable synthesizer. Our results showed that style embeddings extracted from different models outperformed their base representations, and in particular, the Huang style embedding from the CLAP model is markedly superior to the others. To encourage further work, we provide a Python package that implements all our metrics and procedures. We hope this evaluation framework and Python package will encourage advancements in timbre metrics across tasks like generative modeling and instrument retrieval.

\section{Ethics Statements}

This work evaluates models using datasets that primarily feature Western musical instruments, which may reflect a cultural bias toward Western music traditions. We acknowledge this limitation and are enthusiastic about including non-Western musical data in our evaluation framework, as it may both enhance cultural diversity and help reveal biases in the models’ behavior under broader musical contexts.

\section{Acknowledgements}

We thank Mathieu Lagrange for the valuable discussions. This work is supported by the EPSRC UKRI Centre for Doctoral Training in Artificial Intelligence and Music (grant number EP/S022694/1). This research utilized Queen Mary's Apocrita HPC facility, supported by QMUL Research-IT. http://doi.org/10.5281/zenodo.438045.

\bibliography{ref}

\end{document}